\documentclass[aps,preprint,showkeys]{revtex4-1}
\usepackage{graphicx}
\usepackage{amsmath}
\usepackage{makeidx}
\usepackage{amsfonts}
\usepackage{amssymb}
\usepackage{mathtools}
\usepackage{xcolor}

\begin{document}

\title{Entropy production of selfish drivers: Implications for efficiency and predictability of movements in a city}
\date{\today}

\author{Indaco Biazzo}
\affiliation{Politecnico di Torino, Corso Duca degli Abruzzi 24, Torino, Italy}

\author{Mohsen Ghasemi Nezhadhaghighi}
\affiliation{Department of Physics, School of Sciences, Shiraz University, Shiraz 71454, Iran}

\author{Abolfazl Ramezanpour}
\email{aramezanpour@gmail.com}
\affiliation{Department of Physics, School of Sciences, Shiraz University, Shiraz 71454, Iran}
\affiliation{Leiden Academic Centre for Drug Research, Faculty of Mathematics and Natural Sciences, Leiden University, PO Box 9500-2300 RA Leiden, The Netherlands}

\begin{abstract}
Characterizing the efficiency of movements is important for a better management of the cities. More specifically, the connection between the efficiency and uncertainty (entropy) production of a transport process is not established yet. In this study, we consider the movements of selfish drivers from their homes (origins) to work places (destinations) to see how interactions and randomness in the movements affect a measure of efficiency and entropy production (uncertainty in the destination time intervals) in this process. We employ realistic models of population distributions and mobility laws to simulate the movement process, where interactions are modelled by dependence of the local travel times on the local flows. We observe that some level of information (the travel times) sharing enhances a measure of predictability in the process without any coordination. Moreover, the larger cities display smaller efficiencies, for the same model parameters and population density, which limits the size of an efficient city. We find that entropy production is a good order parameter to distinguish the low- and high-congestion phases. In the former phase, the entropy production grows monotonically with the probability of random moves, whereas it displays a minimum in the congested phase; that is randomness in the movements can reduce the uncertainty in the destination time intervals. The findings highlight the role of entropy production in the study of efficiency and predictability of similar processes in a complex system like the city.

\end{abstract}


\maketitle

\section{Introduction}
Human mobility has changed completely over the past two centuries, starting in Britain, in the early nineteenth century, from the dramatic expansion of roads and railways networks \cite{cresswell2006move}.
Studies aiming at describing and understanding the human mobility in space and time started some decades after that \cite{ravenstein1885laws}, and more quantitative and theoretical approaches appear starting from the 1940's \cite{stouffer1940intervening, zipf1946p, adams2001quantitative}.
In recent years, the great improvements of computational resources and data gathering due to the ICT revolution has given new impulses to these studies, for example, the number of physics papers on the arXiv \cite{arxiv} has grown enormously since about 2004-2005 \cite{barbosa2018human}.

The widespread adoption of mobile phones in early 2000's and subsequently of smartphones increased enormously the amount of data available about movements in city. These data fostered the appearance of new researches describing and characterizing urban displacement \cite{gonzalez2008understanding, song2010modelling, predict-sci-2010, Li-nc-2017, Alessandretti2020}.
The private vehicles, until today, cover the majority of trips in large parts of cities \cite{enwiki:1011200149}. There are however multiple drawbacks associated with this situation. Due to the increased private demands of displacements, gridlocks are rising every year \cite{urbanReport, inrixRank, GBTransport}, leading to large economic loss \cite{urbanReport} and increasing urban air pollution \cite{EuropeanEnvironmentAgencyEEA2020}.
In this regards, large investment in public transports infrastructures are mandatory, but also better strategies to understanding and modelling of such complex systems can help to mitigate the undesirable effects. Agent based models have been developed in order to capture the main aspects of the phenomenon, ranging from simple models (see references in \cite{urbanReport}) to very complicated and computationally demanding ones \cite{horni2016multi, kitamura2000micro}. Recently, the pervasive smartphone adoption and the consequent use of navigation Apps have changed the movement habit of individuals, and not always for the better \cite{Apps-ieee-2019}. In this work we want to see how interactions of agents and associated entropy productions influence predictability and efficiency of movements in city. We study a dynamical system with agents moving according to a selfish routing algorithm \cite{efc-srep-2020,  PhysRevResearch.2.032059} and standard models of population and destinations distributions. We consider two kinds of interactions among the agents: spatial and temporal interactions. The former interactions affect the travel times of street segments which is correlated with the flows of agents. In the latter interactions, the travel time information of previous days are taken into account in the selfish routing algorithm. 

We shall see how the predictability and efficiency of the movements depend on the strength of interactions. Two measures of predictability are considered: the entropy production and a distance between distributions of the expected and actual travel times. The measure of efficiency was introduced in \cite{efc-srep-2020} as the inverse of the travel time per the total number of trips. In this study, however, we simulate the movements of many interacting agents, which can also deviate from their optimal trajectories randomly at every time step. We observe that in general the efficiency and predictability of the process are diminished by increasing the interactions in space and time. It is known that sharing the travel information with selfish drivers in the absence of any coordination could lead to traffic "chaos" \cite{Apps-ieee-2019}. We find that very small but nonzero levels of temporal interactions can in fact enhance the predictability by reducing the differences between the expected and actual travel times. Moreover, for large values of spatial interactions, a bit of randomness in the movements results in smaller entropy production by avoiding the congested lines. For the same reason, the efficiency displays a maximum at small values of randomness, when the spatial and temporal interactions are sufficiently strong. 

The paper is structured as follows. In Sec. \ref{S1}, we introduce the models and define the main quantities of the study. In Sec. \ref{S2}, we report the results of numerical simulations and discuss the findings and their consequences. The concluding remarks are given in Sec. \ref{S3}.

\section{Models and Settings}\label{S1}
In this section, we present the main definitions and methods which are used to model the network flow dynamics.   
Consider a city of $N$ sites with local populations $\{m_a:a=1,\dots,N\}$ and total population $M=\sum_a m_a$. The connectivity graph of the city is given by $G(V,E)$ where $V$ is the set of sites and $E$ is the set of directed edges $(ab)$. Here we take a two dimensional square lattice of size $N=L\times L$, where all edges have the same length and the same free travel times. We use the model introduced in Ref. \cite{Li-nc-2017} to produce reasonable population distributions for the model cities. This is a preferential growth model which starts with a seed of population at the center of the lattice. At each step, a site is chosen with probability $p_a \propto (m_a+c_0)$ and a unite of population is added only if there exists a populated site at distance less than $l_0$. In the following, we take $c_0=1$, $l_0=1$, and the population density is fixed to $M/N=10^3$.

Given the population distribution $m_a$, we use the following mobility law \cite{Yan-intf-2014} to construct the flux of movements $m_{a\to b}$ from origins $a$ to destinations $b$, 
\begin{align}\label{mab}
m_{a\to b}=m_ap_{a\to b}=m_a\frac{m_b/M(r_{ab})}{\sum_{c\neq a} m_c/M(r_{ac})},
\end{align}
where $M(r_{ab})$ is the population in the circle of radius $r_{ab}$ centred at site $b$. The ratio $m_b/M(r_{ab})$ can be interpreted as the attractiveness of site $b$ for an individual at site $a$. 

Finally, the flows of movements on edges $(ab)\in E$ are determined by a flux distribution dynamics, as follows.

\begin{figure}
\includegraphics[width=12cm]{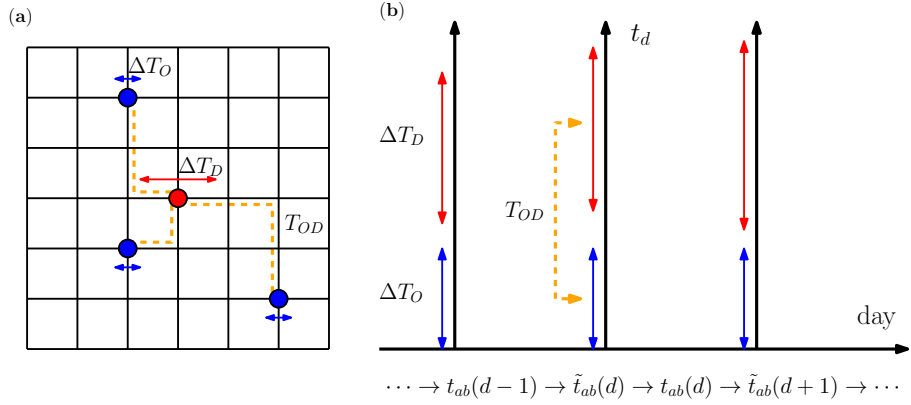} 
\caption{Illustration of the origin to destination trips. (a) In each day, the trips start from the origins in the time interval $\Delta T_O$ (initial state) and reach the destinations in the time interval $\Delta T_D$ (final state) with travel times $T_{OD}$. (b) The actual travel times $t_{ab}(d-1)$ on directed edges $(ab)$ are used to update the expected travel times $\tilde{t}_{ab}(d)$, which are used to find the shortest-time paths in that day.}\label{f1}
\end{figure}

\subsection{The movement process}\label{S11}
Here we describe the process of moving from the origins (Os) to destinations (Ds) in a single day:

\begin{itemize}

\item The starting times of the OD trips are distributed uniformly in the origin time interval $\Delta T_O$. We assume that $\Delta T_O$ is the same for all origins. In each time step, the time increases by $\Delta t=1$. Driver $i$ starts its trip and becomes active at time $t_O(i) \in \Delta T_O$. The trip will become inactive when the driver reaches its destination at time $t_D(i) \in \Delta T_D$. The destination time interval $\Delta T_D$ is determined by the system structure and dynamics (see Fig. \ref{f1}). 
The arrival times of the drivers at destination site $a$ determine the destination time interval of that site $\Delta T_D(a)$. 
The travel time from origin to destination for driver $i$ is denoted by $\Delta t_{OD}(i)=t_D(i)-t_O(i)$. 

\item An active driver $i$ at site $a$ chooses the next site as follows: with probability $\alpha$ the next site is chosen randomly and uniformly from the set of neighbouring sites. With probability $1-\alpha$ the neighbour that minimizes the expected travel time to the destination $D(i)$ is selected. The expected travel time on each directed edge $(ab)$ in day $d$ is denoted by $\tilde{t}_{ab}(d)$. The expected travel times in day $d$ are estimated by using the actual travel times $t_{ab}(d-1)$ from the previous day:
\begin{align}\label{tab0}
\tilde{t}_{ab}(d)=\lambda t_{ab}(d-1)+(1-\lambda) \tilde{t}_{ab}(d-1).
\end{align}
For the initial day $\tilde{t}_{ab}(0)=t_{ab}(0)$, where the $t_{ab}(0)$ are the travel times for free lines. Note that when $\lambda=0$ the expected travel times do not change with day and the drivers always choose the shortest path according to the $t_{ab}(0)$. 

\item Let flow $F_{ab}(t)$ be the number of drivers that enter edge $(ab)$ at time step $t$. Given the flows, the actual travel times are obtained from 
\begin{align}\label{tab}
t_{ab}(F_{ab})=t_{ab}(0)\left(1+g(\frac{F_{ab}}{F_{ab}^*})^{\mu}\right),
\end{align}
with $g (F_{ab}/F_{ab}^*)^{\mu}$ to model the influence of flows on the travel times \cite{BPR-1964,lc-trans-1976,lc-trans-2011}. The actual travel time $t_{ab}$ determines the time that a person spends on edge $(ab)$. Here $F_{ab}^*$ is a measure of the line capacity. For simplicity, we assume that $t_{ba}(0)=t_{ab}(0)$ and $F_{ba}^*= F_{ab}^*$.
In the following, we take $\mu=3$, $t_{ab}(0)=1$, and $F_{ab}^*=M/|E|$ in all the numerical simulations.

\end{itemize}

\begin{figure}
\includegraphics[width=12cm]{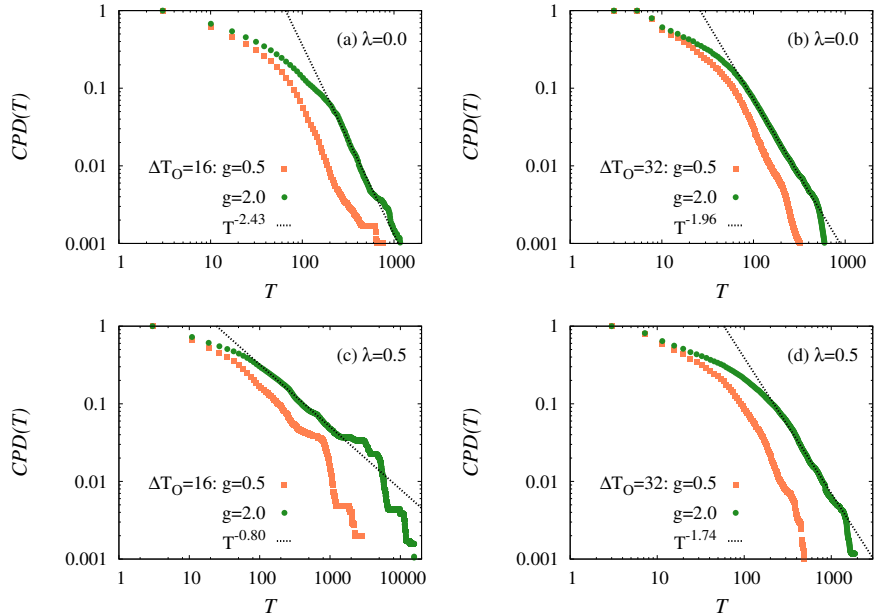} 
\caption{Cumulative probability distribution of the travel times $P(\Delta t_{OD}>T)$ in a single day. The lattice size is $L=40$ and $\alpha=0$.}\label{f2}
\end{figure}

\section{Results and Discussion}\label{S2}
Let us start with the effects of interactions on the cumulative distribution of the travel times $P(\Delta t_{OD}>T)$ in the absence of any randomness in the movements ($\alpha=0$). As Fig. \ref{f2} shows, the travel times increase by introducing the interactions either by considering the effects of flows on the trips (with $g$) or by exploiting the travel information from the previous days (with $\lambda$). We know that interactions can reduce predictability in a system \cite{watts-sci-2006}. Moreover, a selfish way of using the travel information without any coordination could worsen the situation \cite{Apps-ieee-2019}. In the following, we see how the interplay of the above interactions with randomness in the movements affects a measure of predictability of the movement process. In addition, we follow the changes in the efficiency and a measure of entropy production in the system to see how the relation between these quantities depend on the macroscopic state (phase) of the system \cite{efc-srep-2020}.

\subsection{Predictability}\label{S21}
Predictability of the movements can be studied by characterizing the entropy or mutual information of the relevant quantities \cite{predict-sci-2010,predict-jsta-2013,predict-scirep-2013,predict-ieee-2020}. For instance, it is interesting to know how much the travel times depend on the geometrical distances and the starting times of the trips. We also study a time series of the travel times in different days to check a measure of predictability in the stationary state of the process.

For a single stochastic variable one may use the (estimated) entropy to quantify the variable uncertainty \cite{Fano-1961}.
A measure of predictability for two stochastic variables $x,x'$ is provided by the mutual information of the two variables,
\begin{align}
\mathrm{MI}(x,x')=\sum_{x,x'}P(x,x')\log\frac{P(x,x')}{P(x)P(x')},
\end{align}
where $P(x,x'), P(x)$ and $P(x')$ are the joint and marginalized probability distributions of the variables.
Note that large values of mutual information do not necessary mean that accurate prediction is an easy task \cite{predict-profit-2000}.

\begin{figure}
\includegraphics[width=12cm]{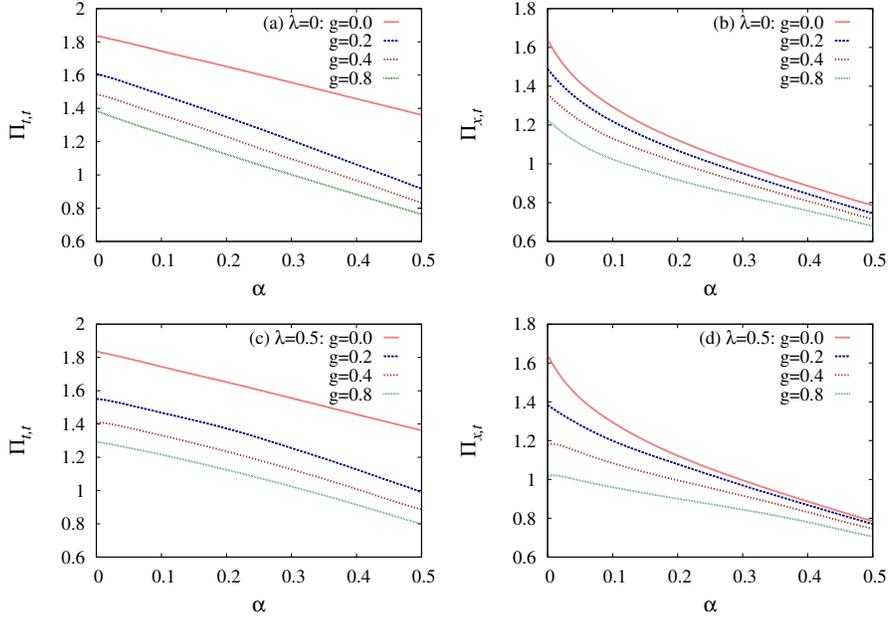} 
\caption{Mutual information of origin times with destination times $\Pi_{t,t}$ and geometrical distances with travel times $\Pi_{x,t}$. The lattice size is $L=40$, and $\Delta T_O=64$. The data are averaged over $200$ independent realizations of the population distribution and the movement process.}\label{f3}
\end{figure}

Let us define the mutual information between the destination and origin times in one day (i.e., along the vertical direction in Fig. \ref{f1}),
\begin{align}
\Pi_{t,t}=\mathrm{MI}(t_O,t_D),
\end{align}
and the mutual information between the travel time and the geometrical distance, 
\begin{align}
\Pi_{x,t}=\mathrm{MI}(\Delta x_{OD},\Delta t_{OD}).
\end{align}
Figure \ref{f3} shows the results of numerical simulations for these quantities in a square lattice of linear size $L=40$.
The above measures of predictability diminish with increasing $g$ and $\alpha$, or by using the travel information from the previous days ($\lambda=0.5$) in the absence of randomness ($\alpha=0$). The difference between the cases $\lambda=0$ and $\lambda=0.5$ gets smaller for larger $\alpha$, where $\Pi_{t,t}$ is even a bit larger in the latter case. Note that when $\lambda=0$ all edges have the same expected travel times $\tilde{t}_{ab}(d)=t_{ab}(0)=1$, where the shortest path is strongly correlated with the geometrical distance $\Delta x_{OD}$. When $\lambda=0.5$, there are some edges with small travel times in the previous day and all drivers are aware of this information which is used to determine the shortest path to their destinations. Therefore, one expects to observe less correlations here between the geometrical distance and the actual travel times. However, the same global information could make the arrival times at the destinations more dependent on the departure times and results to larger mutual information between the two time variables. 

\begin{figure}
\includegraphics[width=12cm]{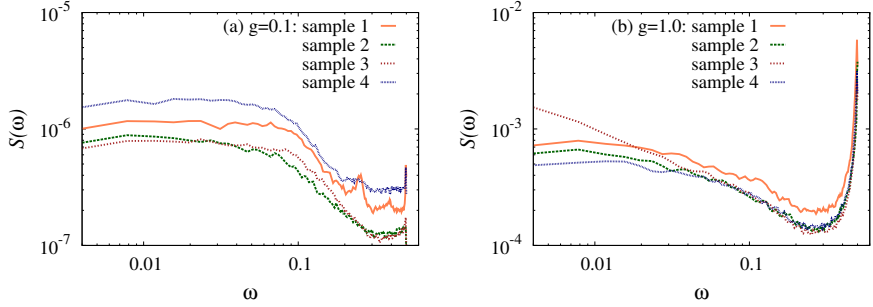} 
\caption{Power spectrum $S(\omega)$ (square of Fourier transform) of $D_d(t,\tilde{t})$ time series in a sequence of $10^5$ days. The lattice size is $L=30$, $\Delta T_O=48$, and $\alpha=0$. The samples show independent realizations of the population distribution and the movement process.}\label{f4}
\end{figure}

\begin{figure}
\includegraphics[width=12cm]{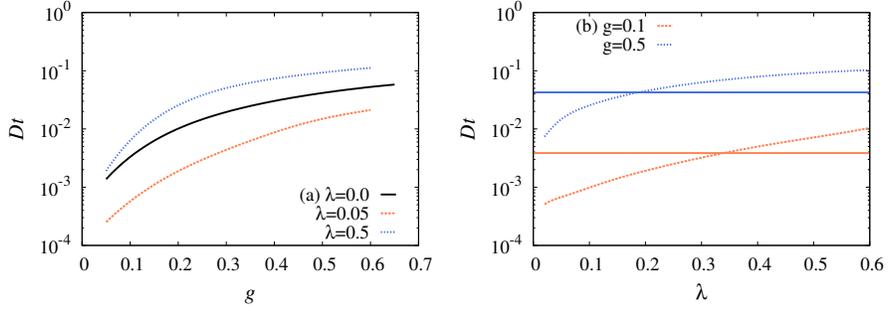} 
\caption{The average deviation from the expected travel times $\langle D_d(t,\tilde{t})\rangle$ in one day. The horizontal lines in panel (b) indicate the values obtained for $\lambda=0$. The lattice size is $L=40$, $\Delta T_O=64$, and $\alpha=0$. The data are averaged over $200$ independent realizations of the population distribution and the movement process in the stationary state. The stationary values are obtained from data samples in an interval of $100$ days after a relaxation stage of $200$ days.}\label{f5}
\end{figure}

Next, we study the changes in the travel times along the links for different days (i.e., along the horizontal direction in Fig. \ref{f1}). Recall that the trips in each day rely on the expected travel times $\tilde{t}_{ab}(d)$ from the previous days. The actual travel times $t_{ab}(d)$ in that day could however be different from the estimations we started with. The relative difference of the two quantities thus provides a measure of unpredictability in the movement process,
\begin{align}
D_d(t,\tilde{t})=\frac{1}{|E|}\sum_{(ab)}\frac{|t_{ab}(d)-\tilde{t}_{ab}(d)|}{\tilde{t}_{ab}(d)}.
\end{align}
Consider a time series of the above quantity as a function of $d$. The power spectrum of such a time series is reported in Fig. \ref{f4}. The appearance of a dominant peak at high frequencies for larger interactions $g$ signals a high level of variation at the shortest time scales (two consecutive days). Figure \ref{f5} shows the stationary values of $D_d(t,\tilde{t})$ versus the model parameters $g$ and $\lambda$. Interestingly, the above quantity exhibits a minimum for very small but nonzero values of $\lambda$ with a discontinuity at $\lambda=0$. This means that some level of information sharing could be helpful even in the absence of any coordination. Moreover, the stronger interactions (larger $g$) result in smaller values and ranges of beneficial $\lambda$, which can be used to reduce the above measure of unpredictability.  On the other hand, for a fixed $\lambda$, the deviations from the expected travel times grow with the strength of interactions as expected.

\subsection{Efficiency and entropy production}\label{S22}
Perhaps the most important quantities of the process regarding its efficiency are the travel times and costs. The average travel time in the process is given by
\begin{align}
\tau_{OD}=\frac{1}{M}\sum_{i} \Delta t_{OD}(i).
\end{align}
The average number of travels per person in a day is obtained from the sum of all the input flows $F_{ab}(t)$ for different edges and times, 
\begin{align}
\sigma_{OD}=\frac{1}{M}\sum_{t} \sum_{(ab)} F_{ab}(t).
\end{align}
This is expected to be proportional to the total energy that is consumed in the process. 
A measure of efficiency then is defined by ratio of the two quantities,
\begin{align}
\eta_{OD}=\frac{1/\tau_{OD}}{\sigma_{OD}}.
\end{align}
This efficiency takes its maximum value for $g=0, \alpha=0$, when both the travel times and the number of edge travels are minimum.     
For each person $i$, we can also define the velocity $v_{OD}(i)=\Delta x_{OD}(i)/\Delta t_{OD}(i)$, given the geometrical origin to destination distance $\Delta x_{OD}(i)$. Then, the average velocity is
\begin{align}
v_{OD}=\frac{1}{M}\sum_{i} v_{OD}(i),
\end{align}
which can be regarded as another measure of the process efficiency \cite{biazzo2019general}.

In the following, we are also interested in the amount of disorder that is generated by the movements. A measure of increase in the system disorder or uncertainty (entropy production) is provided by the relative entropy of the origin and destination time intervals, 
\begin{align}
\Delta S_{OD}=\langle \log \Delta T_D \rangle-\langle \log \Delta T_O \rangle=\frac{1}{N}\sum_a \log \Delta T_D(a)-\log \Delta T_O.
\end{align}
As mentioned before, $\Delta T_O$ is the same for all sites $a$. Other measures of entropy production (or irreversibility) of the process can be defined for example by the relative entropy of the forward (origin to destination) and backward (destination to origin) flows \cite{efc-srep-2020}. In this study, we focus on the above measure $\Delta S_{OD}$, which only needs the destination time intervals at the end of the forward process, to compare with the origin time intervals.

\begin{figure}
\includegraphics[width=12cm]{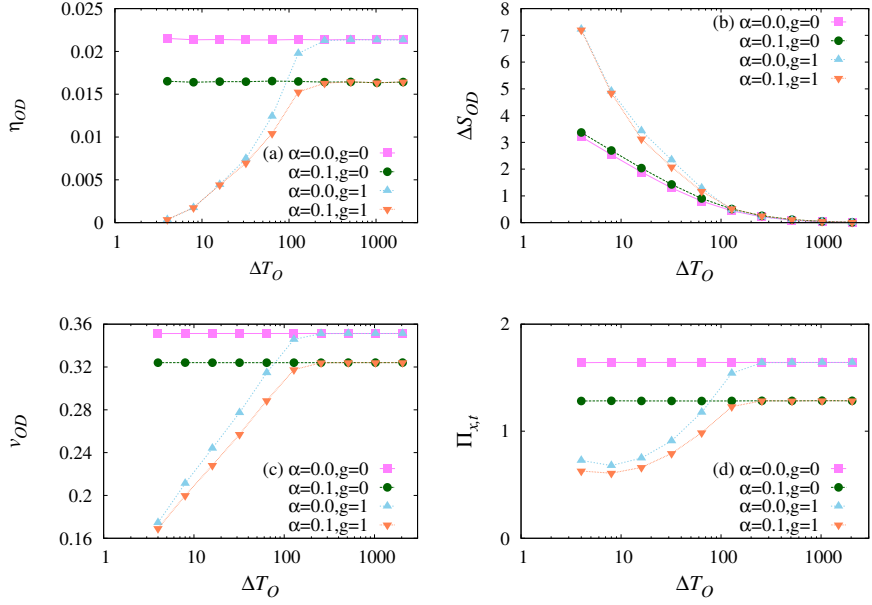} 
\caption{The main quantities vs the origin time interval $\Delta T_O$ for $\lambda=0$. The lattice size is $L=40$. The data are averaged over $200$ independent realizations of the population distribution and the movement process.}\label{f6}
\end{figure}

\begin{figure}
\includegraphics[width=16cm]{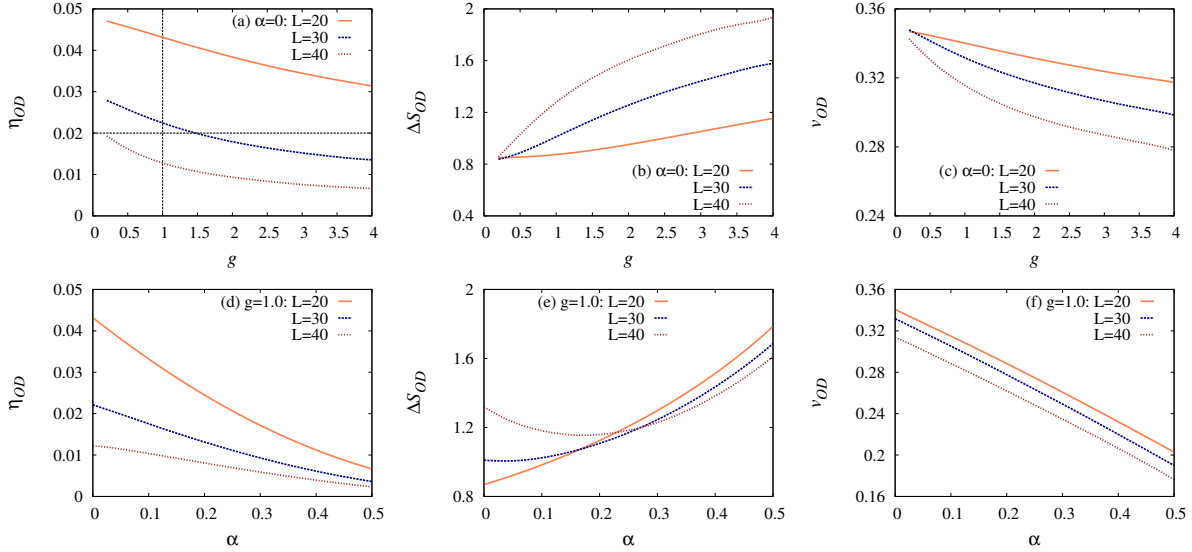} 
\caption{Lattice size effects on the main quantities for $\lambda=0$. Here $\Delta T_O=32, 48, 64$ for $L=20, 30, 40$, respectively to have the ratio $L/\Delta T_O$ fixed. The population density is fixed to $M/N=10^3$. The data are averaged over $1000$ (for $L=20,30$) or $500$ (for $L=40$) independent realizations of the population distribution and the movement process.}\label{f7}
\end{figure}

\begin{figure}
\includegraphics[width=14cm]{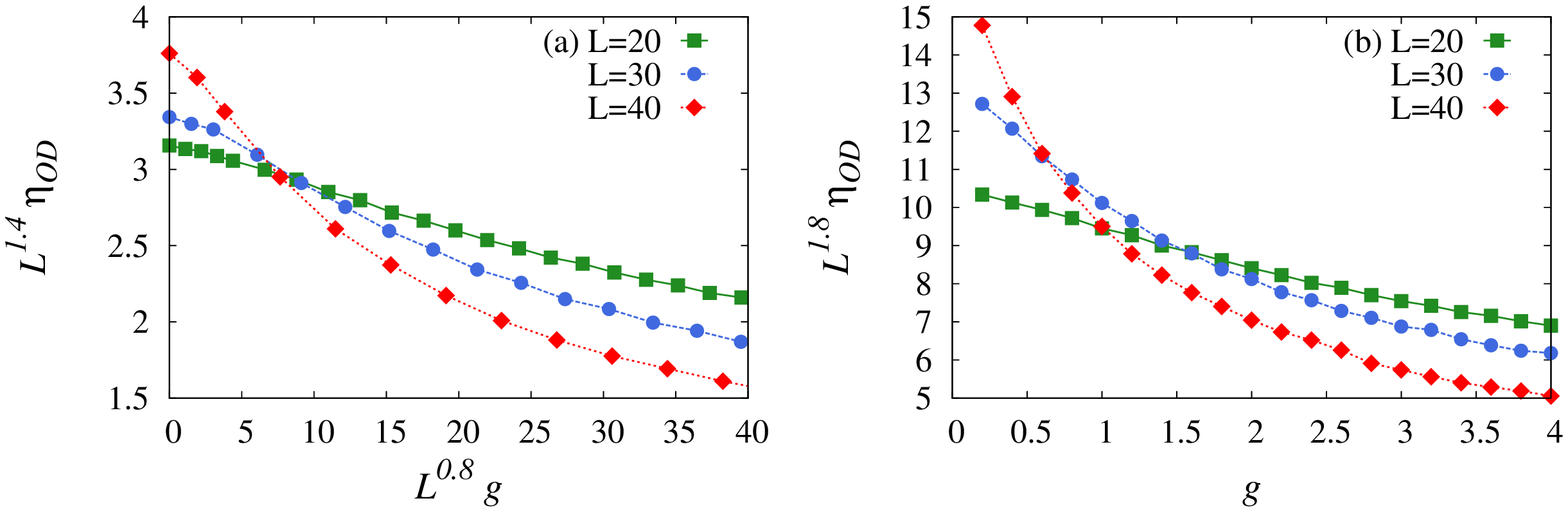} 
\caption{The scaled efficiency vs the (scaled) interaction parameter $g$. Here $\lambda=\alpha=0$, $L=20, 30, 40$ and $\Delta T_O=32, 48, 64$, respectively. The data are averaged over $1000$ (for $L=20,30$) or $500$ (for $L=40$) independent realizations of the population distribution and the movement process.}\label{f8}
\end{figure}

Figure \ref{f6} shows how the above quantities change with the origin time interval $\Delta T_O$ when $\lambda=0$. We see that by increasing $\Delta T_O$ (slower process) the efficiency increases because the interaction parameter $g$ becomes effectively irrelevant for very large time intervals. In this limit, the relative entropy $\Delta S_{OD}$ is nearly zero but it grows continuously as $\Delta T_O$ decreases, separating two phases of high and low efficiency (or velocity).

In Fig. \ref{f7}, we report the lattice size effects when the parameters $g$ and $\alpha$ are varied for $\lambda=0$. We keep the population density $M/N=1000$ and the links' capacity $F^*_{ab}=M/|E|$ for different linear sizes $L=20,30,40$. It is observed that the efficiency and velocity go to zero very rapidly with $L$ even when we increase $\Delta T_O$ to take the ratio $L/\Delta T_O$ fixed. As Fig. \ref{f8} shows, the efficiency scales roughly as $L^{-1.8}$ for $g$ values around $1$. This means that, under the above conditions, the city size is limited by the efficiency or velocity of the movement process. A similar phenomenon is observed in \cite{scaling-sci-2013}, where infrastructure costs put an upper bound on the size of a city (in terms of social interactions). The key point is that here the population is increasing with the size and we have a strongly heterogeneous population distribution concentrated around the center of the lattice. More importantly, it is assumed that link capacities $F^*_{ab}$ are the same in whole the lattice and do not change with the size of the system. These assumptions explain why the efficiency approaches zero by increasing $L$ and why we have to adjust the capacities according to the population size and distribution. The question then is how we should set the local capacities such that the efficiency approaches a desirable limit by increasing the system size.

\begin{figure}
\includegraphics[width=16cm]{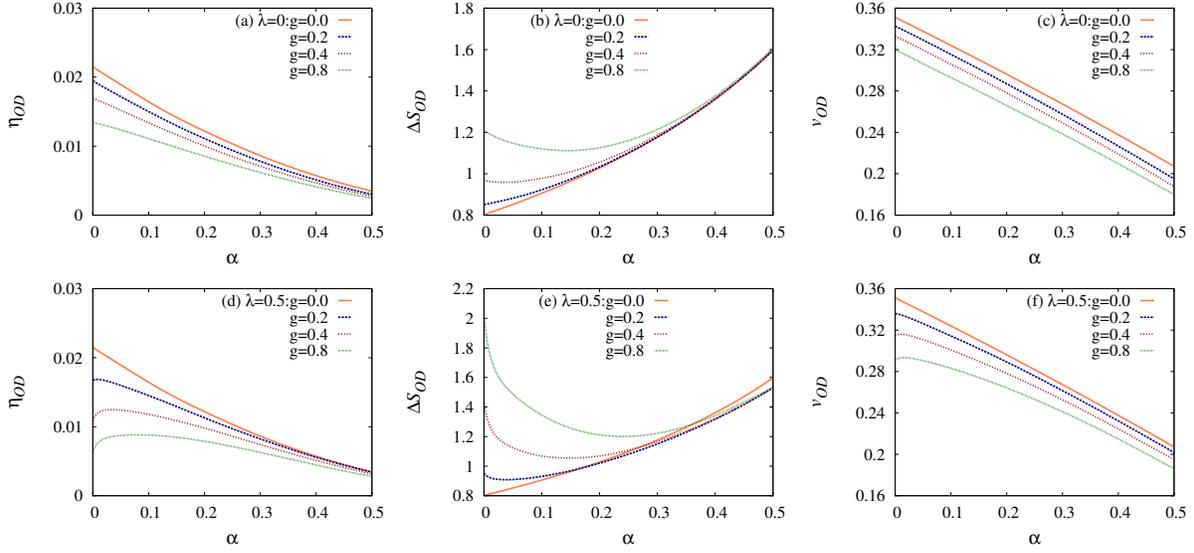} 
\caption{The efficiency $\eta_{OD}$, relative entropy $\Delta S_{OD}$, and velocity $v_{OD}$ vs the randomness parameter $\alpha$ for $\lambda=0$ (top panels) and $\lambda=0.5$ (bottom panels). The lattice size is $L=40$, and $\Delta T_O=64$ here. The data are averaged over $200$ independent realizations of the population distribution and the movement process.}\label{f9}
\end{figure}

Another interesting observation in Fig. \ref{f7} is that $\Delta S_{OD}$ initially decreases with the randomness parameter $\alpha$ and displays a minimum that is more pronounced for larger lattice sizes. This behaviour is better exhibited in Fig. \ref{f9}, where we report the results for a fixed $L=40$ with different values of $g$ and $\lambda$. In fact, the relative entropy increases with $\alpha$ for small values of the interaction parameter. But, for $g$ larger than a critical value, which depends on $L$ and the other model parameters ($\Delta T_O$ and $\lambda$), a bit of randomness in the movements can reduce the relative entropy $\Delta S_{OD}$. That is, random deviations from the shortest paths in this regime allow to have more order in the destination time intervals. When $\lambda=0.5$ this behaviour is observed for a smaller $g$, as using the travel information from the previous days results in a larger effective interaction between the drivers. Besides the minimum displayed by $\Delta S_{OD}$, in Fig. \ref{f9} we also observe a maximum in the efficiency and velocity for $\lambda=0.5$. It means that in this case the randomness is also reducing the total travel times by avoiding the congested links of the network.

\begin{figure}
\includegraphics[width=14cm]{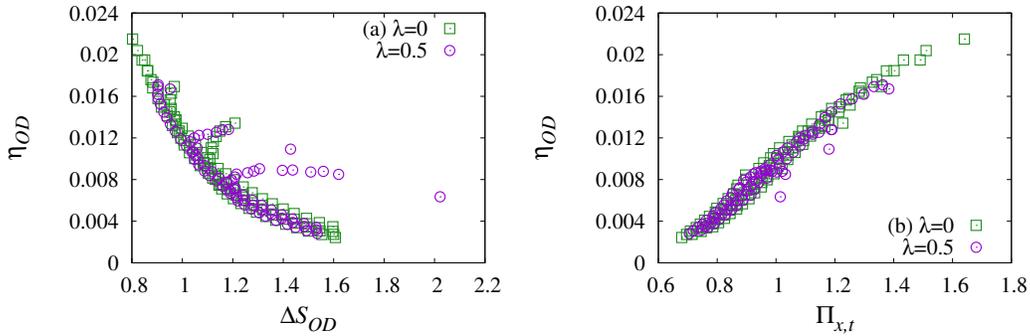} 
\caption{The efficiency vs $\Delta S_{OD}$ and $\Pi_{x,t}$ when both $g$ and $\alpha$ are varied. Here $g=0,0.2,0.4,0.8$, $\alpha \in (0,0.5)$, the lattice size is $L=40$, and $\Delta T_O=64$. The data are averaged over $200$ independent realizations of the population distribution and the movement process.}\label{f10}
\end{figure}

In general, the efficiency is positively correlated with the probability of making a right decision which is expected to increase by reducing the travel uncertainties. This relation should be exhibited by reasonable measures of the two quantities (efficiency and uncertainty), see Fig. \ref{f10}. As the figure shows, there is however a region where both the efficiency and entropy production decrease with increasing $\alpha$. For $\lambda=0$, this happens when randomness in the movements reduces the relative entropy $\Delta S_{OD}$. For $\lambda=0.5$, this region shrinks to a smaller interval, that is from the maximum of efficiency to the minimum of $\Delta S_{OD}$. In other words, a negative correlation between the efficiency and entropy production is better displayed in the more realistic case when the shortest paths are chosen according to the available travel information.

\section{Conclusion}\label{S3}
In summary, numerical simulations of the movement process shows that in general interactions $(g,\lambda)$ increase the travel time, reduce the efficiency, and increase the entropy production. A bit of information sharing by exploiting the travel times from the previous days, could reduce the deviation of the actual travel times from the expected ones, so increasing the process predictability. Moreover, some level of randomness in the movements is beneficial for the efficiency (or inverse of the travel time) of selfish drivers in the congested phase. We also observe a qualitative change in behaviour of the entropy production $\Delta S_{OD}$ with the randomness parameter $\alpha$ as the parameter $g$ increases. In fact, for sufficiently large $g$, one can reduce the uncertainty in the destination time intervals by introducing randomness in the movements. In other words, the response of $\Delta S_{OD}$ to the randomness $\alpha$ can be used to discriminate the low- and high-congestion phases of the flows in the system. 

It would be interesting to see how much these findings are sensitive to details of the movements and definition of the efficiency, predictability, and the entropy production of the process. In this paper, we considered selfish drivers which follow the shortest-time paths and studied small deviations from this routing strategy. Instead, one could start with a large number of interacting random walkers in a random movement process, to investigate the role of an appropriately defined measure of entropy production in the process.

\acknowledgments
This work was performed using the ALICE compute resources provided by Leiden University.

\bibliography{references}

\end{document}